\documentclass[prb,twocolumn,showpacs]{revtex4}% Physical Review B

\usepackage{graphicx}
\usepackage{dcolumn}

\usepackage{eucal}
\usepackage[dvips]{epsfig}
\usepackage{changebar}
\usepackage{amssymb}

\usepackage{amsmath}
\newcommand{\mbf}[1]{\mathbf{#1}}
\newcommand{\f}{\frac}
\newcommand{\muB}{\mu_{\rm{B}}}
\newcommand{\kB}{{\rm k_B}}
\newcommand{\Tc}{T_{\rm{C}}}
\newcommand{\cP}{c_{\rm{P}}}

\newcommand{\EW}[1]{\left\langle{#1}\right\rangle}

\newcommand{\GREEN}[3]{\left\langle\hspace{-.4ex}\left\langle#1;
      #2\right\rangle\hspace{-.4ex}\right\rangle{#3}}

\newfont{\tensy}{cmsy10}
\newcommand{\chemical}[1]{{$\fontdimen16\tensy=3.0pt
                       \fontdimen17\tensy=3.0pt\mathrm{#1}$}}

\begin{document}

%%%%%%%%%%%%%%%%%%%%%%%%%%%%%%%%%%
\newcommand{\csp}{\;,\qquad\qquad} %???
%%%%%%%%%%%%%%%%%%%%%%%%%%%%%%%%%%%%%%%%%%%%%%%%%%%%%%%%%%%%%%%%%%%%%%
\title{Disorder in Diluted Spin Systems}

\author{S. Hilbert}
 \email{hilbert@physik.hu-berlin.de}
\author{W. Nolting}
\affiliation{Institut f\"ur Physik, Humboldt-Universit\"at zu Berlin,
 Newtonstra{\ss}e 15, D-12489 Berlin, Germany}
\date{\today}

\begin{abstract}
The influence of substitutional disorder on the magnetic properties of
diluted Heisenberg spin systems is studied with regard to the magnetic
stability of ferromagnetic diluted semiconductors (DMS). The equation of
motion for the magnon Green's function within Tyablikov
approximation is solved numerically for finite systems. The resulting
spectral density is then used to estimate the magnetization and Curie
temperature of an infinite system. This method is suitable for any form
of a ferromagnetic exchange interaction.
Besides different lattices and spin
magnitude $S$, exchange interactions of different range are examined.
The results show that, for short-range interaction, no magnetic order exists
below the critical percolation concentration, whereas a linear
dependence of the Curie temperature on the concentration of spins is
found for ferromagnetic long-range interaction.

\end{abstract}

\pacs{75.10.Nr, 75.10.Jm, 75.50.Pp, 02.60.-x}

\maketitle

%%%%%%%%%%%%%%%%%%%%%%%%%%%%%%%%%%%%%%%%%%%%%%%%%%%%%%%%%%%%%%%%%%%%%%

\section{Introduction}
\label{sec:Introduction}

The discovery of ferromagnetism in (III,Mn)-V diluted magnetic
semiconductors\cite{OhnoEtal92,OhnoEtal96,ReedEtal01,ParkEtal02}
(DMS) has attracted considerable attention among scientists during the past
years.\cite{Ohno99,DietlEtal00,OhnoMatzukura2001,Dietl02, 
DasSarmaEtal03,PeartonEtal03}
The great interest in these materials is motivated by the idea of
using their spin degrees of freedom in conjunction with their electronic
degrees to build new electronic devices ranging from fast nonvolatile
memories to quantum computers (spintronics).
\cite{ShiEtal96,Ohno98,Prinz98,DiVincenzo99,OhnoEtal00,WolfEtal01,ChibaEtal03}
However, progress has been impeded by the fact
that  most of the DMS studied so far have a Curie temperature $\Tc$ below 
room temperature\cite{ReedEtal01,Dietl02,ChibaEtal02,KuEtal03,ParkEtal02}.

For the development of ferromagnetic DMS with desired properties such as
high Curie temperatures, the theoretical understanding of the magnetism in these
materials plays an important role.
In DMS, a small fraction of the non-magnetic host-semiconductor ions
is substituted by ions, which carry a localized magnetic moment
(spins). These magnetic ions are mainly randomly distributed over the
lattice sites of one of the host-semiconductor species.
This positional disorder breaks the translational symmetry of the
crystal and thus greatly complicates the theoretical
description of the material.

One of the first to consider the substitutional disorder present in
diluted spin systems was Brout in the late 1950s.\cite{Brout59}
Since then, various attempts have been made to tackle the problem of
including the disorder into the theoretical description of
ferromagnetism in diluted spin systems. Many
approaches\cite{Montgomery70,FooWu72,TK72,ElliottPepper73,KumarHarris73,
  HarrisEtal74,Nickel75,Kaneyoshi75,TheumannTK75,LageStinchcombe77,
  TK77,DveyAharonFibich78,
  TK83,KawasakiEtal92,KaneyoshiMascur93,ZhaoHuber96,BouzerarBruno02}
combine mean-field or spin-wave theories with configurational-average
methods, which not only rely on uncontrolled
approximations with respect to the disorder, but also encounter
difficulties when applied to spin systems\cite{Gonis}.
Some use Monte-Carlo simulations
\cite{KleninBlume76,HarrisKirkpatrick77,Klenin79,
  SchliemannEtal01,BerciuBhatt01,CalderonEtal02,  
AlvarezEtal02,BreyGomezsantos03,BouzerarEtal03}, which make the
simplification of classical spins instead of quantum spins.
Other approaches are based on percolation
theory\cite{Griffiths69,LitvinovDugaev01,KaminskiDasSarma03} or replica
methods\cite{SherringtonSouthern75,EggenkampEtal95}, but these treat the
magnetism itself only on a mean-field level. However, one must properly
take into account the positional disorder of the spins and their quantum
nature to obtain reliable predictions about the magnetic properties of
diluted spin
systems.\cite{BerciuBhatt01,BouzerarEtal03,Schliemann03,Timm03}

In this article, we will present an approach, which is
based on the Tyablikov approximation\cite{BogolyubovTyablikov59} to the
Heisenberg model and uses numerical studies of finite systems to
estimate spontaneous magnetizations and Curie temperatures of the
corresponding systems in the thermodynamic limit.
It should be mentioned here that the idea of using numerical studies of finite
spin systems in spin-wave-type approximation is not 
new.\cite{Huber74,AlbenEtal77,BerciuBhatt02} According to our knowledge,
however, the obtained 
spectral densities have never been used before to calculate
spontaneous magnetization and Curie temperatures.
The presented approach is suitable for any form of
ferromagnetic exchange interaction. Within this approach, we are able to
treat the quantum fluctuations of the spins within random-phase
approximation. It means that our approach goes beyond the
classical-spin approximation and the mean-field
theory, which notoriously overestimates ferromagnetic
stability. Furthermore, we are able to treat the positional disorder of
the spins numerically exact in contrast to approaches employing
configurational-average methods.

This article is organized as follows: In section \ref{sec:Model},
we will generalize the Tyablikov approximation to spin systems without
translational symmetry.
Section \ref{sec:Numerical_Studies} is concerned with the numerical
studies we performed on the basis of the generalized Tyablikov approximation.
In subsection \ref{subsec:Algorithm}, we will explain the algorithm we
used for our numerical studies. In the following subsections, we will discuss the 
results we obtained for systems with ferromagnetic short-range,
ferromagnetic long-range and oscillating long-range exchange interactions.
In section \ref{sec:Summary}, we will conclude the article with a summary and
proposals for future work.

\section{The Model}
\label{sec:Model}

To study the magnetic properties of a material with a
concentration $c\in[0,1]$ of atoms carrying a localized magnetic moment, we
consider a lattice with a fraction $c$ of the lattice sites
occupied by a spin. We assume the dynamics of these spins to be
described by the isotropic Heisenberg Hamiltonian
\begin{equation}
\label{eq:Heisenberg_Hamiltonian_1}
  H=-{\sum_{i,j=1}^{N}}J_{ij}\, \mbf{S}_i\cdot \mbf{S}_j
-\f{1}{\hbar}g_J \muB B {\sum_{i=1}^{N}} S_i^z\;.
\end{equation}
Here, $i$ and $j$ label the occupied lattice sites and
\mbox{$\mbf{S}_i=\left(S_i^x,S_i^y,S_i^z\right)$} 
is the spin operator of the localized magnetic moment at lattice site
$i$ with lattice vector $\mbf{R}_i$. The spins -- whose total number
is denoted by $N$ -- interact via a Heisenberg
exchange coupling with exchange parameters $J_{ij}$ obeying
\mbox{$J_{ij}=J(|\mbf{R_i}-\mbf{R}_j|)$} and \mbox{$J_{ii}=0$}. In
addition, the Hamiltonian contains a Zeeman coupling of the spins to a
homogeneous  external magnetic field \mbox{$\mbf{B}=(0,0,B)$}. 

The Tyablikov approximation\cite{BogolyubovTyablikov59} has been developed
for and very successfully applied to concentrated spin systems with
translational symmetry. 
In the following, we will generalize the Tyablikov approximation
to derive expressions for the magnetization and the Curie temperature
for diluted spin systems, where we can not presuppose translational
symmetry.
Thereby, we will keep the form of the appearing expressions as close as
possible to those known from the usual Tyablikov approximation. However, 
the absence of translational symmetry complicates the evaluation of
the derived expressions, as will be seen later.

We will first consider spins with
$S=1/2$. In this case, the magnetic properties of the system can be
studied by use of the retarded magnon Green's function 
\begin{equation}
\label{eq:Green_Funktion_Df}
  G_{ij}(E)=\GREEN{S_i^+}{S_j^-}{_E^{ret}}\;
\end{equation} 
containing the step operators \mbox{$S_i^\pm=S_i^x\pm i S_i^y$}.
The equation of motion for $G_{ij}(E)$ reads
\begin{multline}
\label{eq:Bewegungs_Gleichung_voll}
  \left(E-g_J \muB B\right)G_{ij}(E) = 
  2\hbar^2\delta_{ij}\EW{S_i^z}
   + 2\hbar\sum_{m}J_{im} \times \\
\times \left( 
  \GREEN{S_i^+ S_m^z}{S_j^-}{_E^{ret}}
  -\GREEN{S_m^+ S_i^z}{S_j^-}{_E^{ret}}\right)\;.
\end{multline}
Making the Tyablikov approximation, which consists in decoupling the
higher Green's function on the rhs. of
\eqref{eq:Bewegungs_Gleichung_voll}, and assuming a uniform magnetization
\mbox{$\EW{\mbf{S}_i}\equiv(0,0,\EW{S^z})$}, one obtains after rearrangement:
\begin{equation}
\label{eq:Bewegungs_Gleichung_Tyablikov}
  \sum_{m} \left[\left(
  \frac{E-g_J \muB B}{2\hbar\EW{S^z}}
  -\sum_{n} J_{in}\right)\delta_{im}+ J_{im}
  \right]G_{mj}=\hbar\delta_{ij}\;.
\end{equation}
In the following, it is convenient to interpret this equation as a
matrix equation:
\begin{equation}\label{eq:Bewegungs_Gleichung_Matrix}
  \left(\frac{E-g_J \muB B}{2\hbar\EW{S^z}}\mbf{1}-\mbf{H}
\right)\mbf{G}(E)=\hbar \mbf{1}\;.
\end{equation}
Here, the Hamilton matrix $\mbf{H}$, which is defined by its matrix elements
\mbox{$H_{ij}=\delta_{ij}\sum_{n=1}^{N}J_{in}-J_{ij}$}, has been
introduced.

In order to solve equation \eqref{eq:Bewegungs_Gleichung_Matrix} for
$\mbf{G}$, one has to diagonalize $\mbf{H}$.
In the concentrated case, this can be done by a Fourier transformation
because of the translational symmetry of the system. For a diluted system,
translational symmetry is absent, and hence, no general method exists for
this task. However, for a finite system the Hamilton
matrix can be diagonalized numerically.
Since $\mbf{H}$ is
Hermitian, this diagonalization yields $N$ real eigenvalues $E_r$ with
eigenvectors, which can be chosen to form an orthonormal basis. For
notational convenience, we will use indices $r$ and $r'$ to refer to
matrix elements in this '$\mbf{H}$-diagonal' basis, whereas we retain
$i$ and $j$ to refer to matrix elements in the 'site-diagonal' basis,
where matrix indices are identified with site labels.

After going to the $\mbf{H}$-diagonal basis, one obtains for the magnon
Green's function: 
\begin{equation}
 G_{rr'}(E)= \frac{2\hbar^2\EW{S^z}}
 {E-g_J \muB B - 2\hbar\EW{S^z}E_{r}+i0^+}\delta_{rr'}\;.
\end{equation}
From this expression, one can read off  the magnon spectral density
\mbox{$S_{rr'}(E)=-\Im \left( G_{rr'}(E)\right)/\pi$}:
\begin{multline}
  S_{rr'}(E)=-\f{1}{\pi}\Im \left( G_{rr'}(E)\right)\\
= 2\hbar^2\EW{S^z}
  \delta\left(E-g_J \muB B- 2\hbar\EW{S^z}E_r\right)\delta_{rr'}\;.
\end{multline}
Thus, the energies \mbox{$\bar{E}_r=g_J \muB B- 2\hbar\EW{S^z}E_r$} are
the excitation energies of the elementary excitations of the systems,
which we call magnons.

To obtain the local magnon spectral density $S_{ii}(E)$, one has to
transform  $S_{rr'}(E)$ back to the site-diagonal basis. However, this may
yield different $S_{ii}(E)$ for different lattice sites $i$. This
contradicts the assumption of a uniform magnetization, since
$S_{ii}(E)$ and $\EW{S_i^z}$ are connected by the spectral theorem:
\begin{equation}
  \EW{S_i^z}=\hbar S-\frac{1}{\hbar}\EW{S_i^-S_i^+}=
\hbar S-\frac{1}{\hbar^2}
\int_{-\infty}^{\infty}\frac{S_{ii}(E)}{e^{\beta E}-1}dE\;, 
\end{equation}
where \mbox{$\beta=1/\kB T$} with $\kB$ as the Boltzmann constant and
$T$ as the temperature.
One can circumvent the problem by abandoning the
simplification of a uniform magnetization at the cost of having
to solve a system of $N$ non-algebraic equations in an iterative
procedure, where  one has to diagonalize an $N\times N$
matrix similar to $\mbf{H}$ in each step.
Here, a numerically less
demanding approach is taken, which we believe to capture the
important physics of the system: We consider $\EW{S^z}$ an
approximation for $\EW{S_i^z}$ and use the site-averaged spectral
density 
\begin{equation}
\label{eq:Spektraldichte_Df}
  \tilde{S}_{ii}(E)=\f{1}{N}\sum_{i=1}^{N}S_{ii}(E)
  =\f{1}{N}\sum_{r=1}^{N}S_{rr}(E)
\end{equation}
as an approximation for $S_{ii}(E)$ and then use the spectral theorem:
\begin{multline}
  \EW{S^z}\approx
  \EW{S_i^z}=\hbar S-\frac{1}{\hbar^2}
\int_{-\infty}^{\infty}\frac{S_{ii}(E)}{e^{\beta E}-1}dE \\
\approx\hbar S-\frac{1}{\hbar^2}
\int_{-\infty}^{\infty}\frac{\tilde{S}_{ii}(E)}{e^{\beta E}-1}dE\;.
\end{multline}
In short, we have to solve:
\begin{equation}
  \EW{S^z}=
\hbar S-\frac{1}{\hbar^2}
\int_{-\infty}^{\infty}\frac{\tilde{S}_{ii}(E)}{e^{\beta E}-1}dE\;.
\end{equation}
After inserting \eqref{eq:Spektraldichte_Df} and doing some
rearrangements, this yields:
\begin{equation}
\label{eq:Sz_formula}
 \EW{S^z}=\f{\hbar S}{1+\f{2}{N}\sum_{r=1}^{N}
\left[e^{\beta\left(g_J \muB B+2\hbar\EW{S^z}E_r\right)}-1\right]^{-1}}\;.
\end{equation}
From this equation, one obtains an expression for the Curie temperature by
linearization with respect to $\EW{S^z}$\cite{Nolting7}:
\begin{equation}
\label{eq:Tc_formula}
   \Tc=\f{\hbar^2}{2 \kB}\left(\frac{1}{N} \sum_{r=1}^{N}\frac{1}
  {E_r} \right)^{-1}\;.
\end{equation}
Note that for translationally invariant systems, the indices $r$ can 
be identified with the wave vectors of the first Brillouin zone, and
thus the expressions \eqref{eq:Sz_formula} and \eqref{eq:Tc_formula}
reduce to the standard Tyablikov formulae for concentrated spin systems.

For general $S$-values, an analogous derivation using a set of Green's
functions\cite{TK62,Praveczki63,Callen63,NoltingMagnetismus2} instead of
the Green's function \eqref{eq:Green_Funktion_Df} alone leads to the
implicit equation
\begin{equation}
  \label{eq:Sz_all_S}
  \EW{S^z}=\hbar\frac
  {\left(1+S+\Phi\right)\Phi^{2S+1}+
    \left(S-\Phi\right)\left(1+\Phi\right)^{2S+1}}
  {\left(1+\Phi\right)^{2S+1}-\Phi^{2S+1}}
\end{equation}
for the magnetization. Here, the average magnon number
\begin{equation}
  \label{eq:Magnon_number_Df}
  \Phi=\frac{1}{N}\sum_r\frac{1}{e^{\beta\left(g_J \muB
        B+2\hbar\EW{S^z}E_r\right)}-1}
\end{equation}
has been used as a convenient abbreviation. The Curie temperature is then
given by
\begin{equation}
  \label{eq:Tc_all_S}
  {\kB} \Tc=\frac{2}{3}\hbar^2 S\left(S+1\right)
  \left(\frac{1}{N}\sum_r\frac{1}{E_r} \right)^{-1}
\end{equation}
for general $S$-values.

In the following we will only consider systems in an infinitesimal 
external magnetic field $B=0^+$ just to break rotational invariance and
therefore, usually drop terms including the $B$-field explicitly. 

\section{Numerical Studies of Finite Systems}
\label{sec:Numerical_Studies}

\subsection{Algorithm}
\label{subsec:Algorithm}

The equations \eqref{eq:Sz_formula} to \eqref{eq:Tc_all_S} are of no
practical use for the calculations of
the magnetic properties of the system unless one can compute the
eigenvalues $E_r$ of the Hamilton matrix $\mbf{H}$. As already mentioned
above, the Fourier transformation, which is the weapon of choice for
concentrated spin systems, does not work for diluted systems,
because translational symmetry is absent. However, for a finite
system, we can diagonalize the Hamilton matrix numerically to obtain its
eigenvalues. 
Compared to configurational average methods,
the numerical diagonalization has the advantage that no approximations
with respect to the disorder are needed.
Furthermore, the spins can be treated quantum mechanically and the
simplification to classical spins as in usual Monte Carlo simulations is
not necessary. 

The algorithm we used in our
numerical studies of finite systems is as follows: First of all, 
the system parameters must be fixed. These are the concentration $c$ of
spins on the lattice, the exchange parameter $J\left(R\right)$ as a
function of the distance $R$ between two spins and the size and geometry
of the system.
The next step is to decide by the help of a pseudo-random-number
generator which lattice sites of the system are
occupied by a spin. Now that the positions of all occupied lattice sites
in the system are known, the exchange parameters
\mbox{$J_{ij}=J\left(\left|\mbf{R}_i-\mbf{R}_j\right|\right)$} between
occupied lattice sites $i$ and $j$ can be calculated, and hence,
the matrix elements $H_{ij}$. Now, the matrix $\mbf{H}$ can be
diagonalized numerically using standard diagonalization routines for large
matrices.

For the shape of the simulated samples, we chose a cuboid lattice section,
which we closed to a torus to eliminate surface effects. To avoid
multiple- and self-interaction due to the periodic boundary
conditions, we cut off any long-range interaction at a distance half the
system size.
We chose the size of the samples such that for each concentration we
obtained approximately the maximal number of spins in the system that we
could handle (ca. $14\,000$ spins on an AMD Athlon PC, 2GHz CPU, 1 GB
RAM).
For simplicity, we only considered the case that each lattice site is
occupied by a spin with probability $c$ independent of the occupation of
neighboring sites. However, if one is interested in clustering effects, a
different routine to distribute the spins on the lattice could be used.

To decrease statistical errors, we averaged the eigenvalue distribution,
i.e. $\tilde{S}_{ii}(E)$,
over several simulated systems with equal system parameters, but
different pseudo-random distribution of the spins. Test calculations
showed that this averaging seems hardly necessary for the large samples
we used, which is due to the self-averaging property of the
site-averaged local magnon spectral density
$\tilde{S}_{ii}(E)$\cite{Gonis}. 
As a trick to smoothen the eigenvalue spectra, we sometimes used
slightly different sample geometries for different simulation runs and
weighted the obtained eigenvalue distributions with their number of
lattice sites\cite{AlbenEtal77}. Further weighting was not necessary,
since for a fixed system size the probability of generating a certain
distribution of spins with the used algorithm coincides with the
statistical weight of this distribution among those having an equal number
of lattice sites.\cite{Binder79}

The disadvantage of the numerical diagonalization is that only finite
systems can be treated, but there is no spontaneous
magnetization in the isotropic Heisenberg model for any finite
system. \cite{Bogoliubov60,Bogoliubov62}
The Tyablikov approximation preserves this property, which is 
seen in our calculations by the fact that, due to its structure,
$\mbf{H}$ has at least one zero-eigenvalue, which has a finite spectral
weight for finite systems. More precisely, the number $N_{\rm zeros}$ of
zero-eigenvalues of $\mbf{H}$ equals the number $N_{\rm clusters}$
of clusters of spins connected 
by a non-zero exchange interaction for systems with only ferromagnetic
interaction and $N_{\rm zeros}\geq N_{\rm clusters}$ for systems containing
antiferromagnetic interaction.\cite{Hilbert04}
Hence, \eqref{eq:Sz_formula} and
\eqref{eq:Tc_formula} always yield zero for the spontaneous
magnetization and the Curie temperature, if the computed
eigenvalues of the simulated samples are used directly as input.
In order to obtain a finite spontaneous magnetization and Curie
temperature, one has to go to infinite systems. However, the eigenvalue
distribution of a sufficiently large finite system represents a good
approximation to the eigenvalue distribution of the corresponding
infinite system except for energies close to zero, where the
discreteness of the eigenvalue spectrum of the finite system comes into
play. We solved this problem by shifting the
zero-eigenvalues of $\mbf{H}$ slightly to higher energies, which will be
explained in more detail in the following subsection.

%%%%%%%%%%%%%%%%%%%%%%%%%%%%%%%%%%%%%%%%%%%%%%%%%%%%%%%

\subsection{Short-Range Interaction}
\label{subsec:scNN}

As first application we investigate systems of spins on a simple
cubic (sc) lattice with nearest-neighbor interaction only:
\begin{equation}
  J\left(R\right)=\begin{cases}
    J^0 &\text{for } R =a,\\
    0   &\text{otherwise.}\\
  \end{cases}
\end{equation}
Here, $a$ denotes the nearest neighbor distance and $J^0$ the nearest
neighbor interaction strength.

\begin{figure}[tb]
\centerline{\includegraphics[width=1\linewidth]
{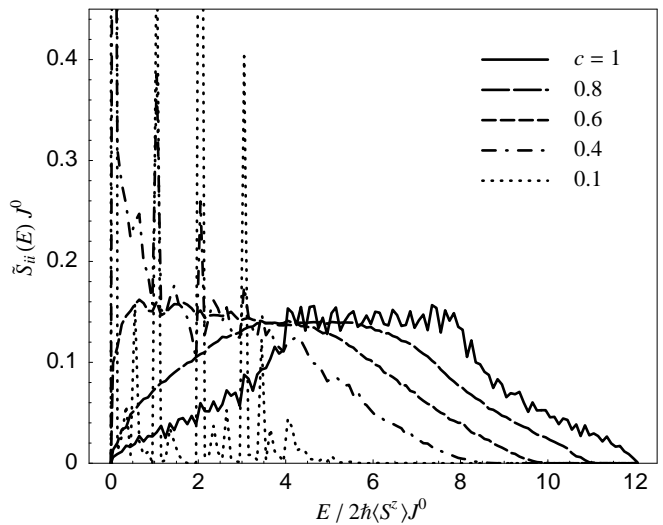}}
\caption{
\label{fig:scNN_Sii}
  Site-averaged local magnon spectral density $\tilde{S}_{ii}(E)$
for nearest-neighbor interaction on an sc lattice for various
concentrations $c$ of spins (shown as histogram of the eigenvalue
distribution of the Hamilton  matrix, each curve is an average of 5 to
10 configurations of $8\,000$ to $14\,000$ spins)
} 
\end{figure}

In figure \ref{fig:scNN_Sii}, we plotted
the spectral density for different concentrations $c$ of spins. The
spectral density is zero for all $E<0$ for all concentrations, which is
due to the fact that $\mbf{H}$ is positive semi-definite for systems
with purely ferromagnetic interactions. The calculations for the
concentrated system show the well-known symmetric shape of a simple
cubic density of states. Dilution increases the spectral density for
lower energies at the cost of the spectral density at higher
energies. For concentration above $c\approx0.5$, the spectral 
density stays smooth, whereas for lower concentrations, peaks
at certain energies
\mbox{($E/2\hbar\EW{S^z}J^0=0,1,2,2-\sqrt{2},\ldots$)} become apparent,
which constitute the whole spectral density for $c<0.3$.
The smooth part of the spectral density can be attributed to one large
percolating cluster of spins, which 
has a quasi-continuous spectrum. This percolating cluster only exists
above the critical percolation concentration
$\cP$, which is \mbox{$\cP^{\rm sc}\approx0.3$} for the simple cubic lattice
with nearest-neighbor interaction.\cite{Kirkpatrick73,Stauffer95}
The peaks stem from smaller clusters, which contain a
considerable fraction of the spins only for concentrations around and
below $\cP$ and which have a discrete spectrum. 

\begin{figure}[tb]
\centerline{\includegraphics[width=1.0\linewidth]
{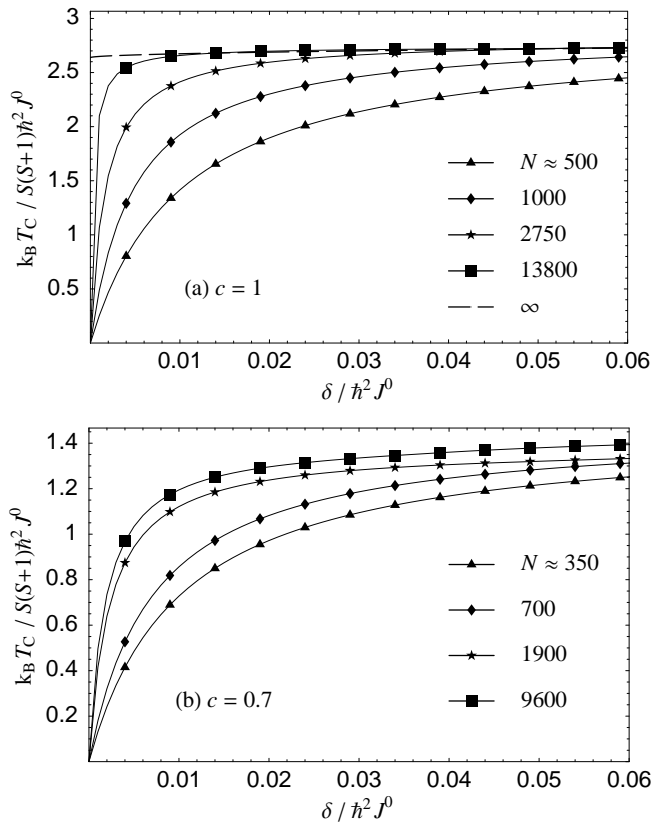}}
\caption{
\label{fig:scNN_Tc_delta}
Curie temperature $\Tc$ as function of the energy shift $\delta$ used
for the low-lying eigenvalues of $\mbf{H}$ for different number $N$ of spins
in the system for (a) $c=1$ and (b) $c=0.7$
}
\end{figure}

To obtain non-zero Curie temperatures from the obtained
$\mbf{H}$-eigenvalues of finite systems, we used a small energy $\delta$
to shift the eigenvalues close to zero slightly to higher energies. To
be more specific, we used
\begin{equation}
  \label{eq:eigenvalue_trafo}
  E_r\mapsto\begin{cases}
\delta&\text{for }E_r<\delta,\\
E_r& \text{otherwise}
  \end{cases}\hspace{1em}
\forall r\in\left\{1,\ldots,N\right\}
\end{equation}
for practical reasons.
The effect of this shift $\delta$ on the Curie temperature $\Tc$ is
illustrated in figure \ref{fig:scNN_Tc_delta}. Let us first discuss the
curves for $c=1$, where the 
numerically obtained data for finite systems can be compared with the
analytical results for the infinite system. For $N=\infty$, the Curie
temperature has a finite value at $\delta=0$ and is quite insensitive to
small shifts  $\delta\ll J^0$.
For large enough but finite $N$, the calculated Curie temperature
rapidly rises from $\Tc=0$ at $\delta=0$ to a value comparable to the
infinite system at $\delta\approx J^0/N$. In this steep region,
$\Tc(\delta)$ is dominated by the shifted zero-eigenvalue, which has a
spectral weight $w_{\rm zeros}=N^{-1}$, and
$\Tc(\delta)\approx \frac{2}{3}\hbar^2 S\left(S+1\right) N \delta$.
For $\delta\gg J^0/N$, however, the
shifted zero-eigenvalue becomes less important, and the curves of the
finite system and the infinite system are almost identical. Hence, the
Curie temperature for the finite system with $c=1$, large $N$ and
$J^0/N\ll\delta\ll J^0$ approximates the Curie temperature for the
infinite system with $c=1$ very well. Furthermore, for large $N$ and
$J^0/N\ll\delta\ll J^0$, the results are quite insensitive to the actual
choice of $\delta$.

For $c<1$, the situation is more complicated. Firstly, there are no
exact analytical results for $\Tc$ we could compare our numerical
results with. Furthermore, for an exchange interaction with finite
range, there is always a finite fraction of spins in smaller clusters.
Hence, always a finite fraction $w_{\rm zeros}=N_{\rm zeros}/N$ of
eigenvalues is zero. Therefore,
$\Tc(\delta)$ maintains a finite slope $\propto w_{\rm zeros}$ for
$\delta\ll w_{\rm zeros} J^0$ even for very large $N$ as is seen in
figure\ref{fig:scNN_Tc_delta}(b).
This problem could be avoided if only the percolating cluster, which is
the only cluster that can support long-range order in the system, were
considered in the calculation of $\Tc$. However,
identifying and removing all the eigenvalues 
due to smaller clusters is numerically quite demanding.
Moreover, for concentrations well above the critical
percolation concentration, the number of smaller clusters is very
small\cite{Stauffer95} and thus $w_{\rm zeros}\ll1$.
Hence, the behavior of $\Tc(\delta)$ is dominated by the
eigenvalues of the percolating cluster for $w_{\rm zeros}J^0\ll\delta$
and $c>\cP$
and is not very sensitive to changes in $\delta$ for $\delta \ll J^0$ as seen
in figure \ref{fig:scNN_Tc_delta}(b).

For $c\lesssim \cP$ most or all 
eigenvalues are due to smaller clusters and therefore, the applicability
of the presented method may be questioned. 
However, since a large fraction of the eigenvalues is zero,
$\Tc(\delta)\approx \frac{2}{3}\hbar^2 S\left(S+1\right) \delta$, which 
simply means that $\Tc\approx0$ for small $\delta$. This agrees with
percolation theory\cite{Stauffer95}, from which follows that $\Tc=0$ for
$c< c_{\rm P}$, since there can not exist any long-range ordering in the
system. Therefore, we believe that the Curie temperature calculated
using the eigenvalues of a large finite 
system and a shift $\delta$ chosen such that $J^0/N\ll\delta\ll J^0$
gives a good estimate for 
the Curie temperature of the corresponding infinite system for finite
dilution, too.
In the following, all given values for the Curie temperature and spontaneous
magnetization refer to $\delta=0.01J^0$, which
corresponds for systems with 
$N\approx 14\,000$ to a Curie temperature $\Tc(\delta)$ just outside the
steep region at small $\delta$.

\begin{figure}[tb]
\centerline{\includegraphics[width=1\linewidth]
{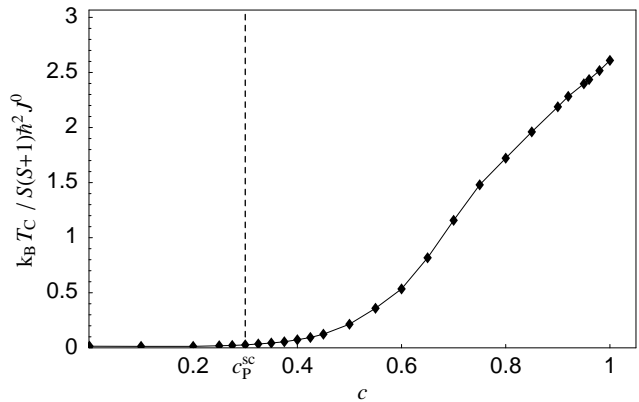}}
\caption{
\label{fig:scNN_Tc_c}
Curie temperature $\Tc$ as function of the concentration $c$ 
for nearest-neighbor interaction on an sc lattice
} 
\end{figure}

\begin{figure}[tb]
\centerline{\includegraphics[width=1\linewidth]
{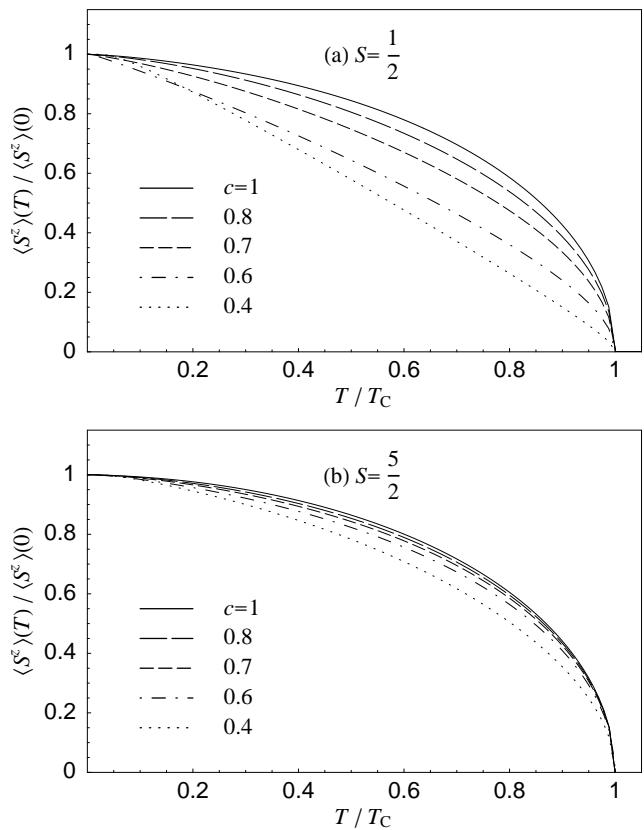}}
\caption{
\label{fig:scNN_Sz_red}
Reduced magnetization $\EW{S^z}/\EW{S^z}(T=0) $ as function of
reduced temperature $T / \Tc$ for  nearest-neighbor
interaction on an sc lattice for various concentrations $c$
for (a) $S=1/2$ and (b) $S=5/2$
} 
\end{figure}

The Curie temperature $\Tc$ we calculated for various concentrations is 
shown in figure \ref{fig:scNN_Tc_c}. Below $c=0.3$, the Curie temperature is
practically zero in accordance with percolation
theory. Above $c=0.3$, the Curie temperature increases with increasing
concentration and reaches its maximum value \mbox{$\kB \Tc=2.61
  S(S+1)\hbar^2 J^0$} at $c=1$ in good agreement with the analytical
result of \mbox{$\kB \Tc=2.64 S(S+1)\hbar^2 J^0$}. 

The question of the effect of the substitutional disorder at finite
dilution on the Curie temperature is connected to the question, how one
defines the ordered system for finite dilution. Placing the
remaining spins of the diluted system on an artificial lattice with an
appropriate lattice constant to regain translational invariance would
give a Curie temperature merely depending on how one alters the
nearest-neighbor exchange interaction strength with increasing lattice
constant. The virtual-crystal approximation, which is widely used
to incorporate dilution effects because of its simplicity, would give
$\Tc(c)=c\,\Tc(1)$, which is certainly wrong for 
low $c$ in the case of an exchange interaction with finite range.

The reduced magnetization as a function of
temperature is shown in figure \ref{fig:scNN_Sz_red} for several
concentrations. For lower concentrations, the magnetization curves show
a more linear decrease with temperature. This effect is less pronounced
for higher spins. Any anomalous curvature can not be found.

\subsection{Ferromagnetic Long-Range Interaction}
\label{subsec:FM_Long_Range}

The experimental fact of Curie temperatures of more than 100 K in
DMS such as \chemical{Ga_{1-{\it x}}Mn_{\it x}As}
with concentrations as low as $\approx0.05$ of
magnetic atoms \cite{OhnoEtal96,MatzukuraEtal1998,BeschotenEtal1999} implies 
that the exchange interaction between the localized moments is not only
quite strong, but must be very long-ranged in these materials.
To study the influence of the range of a ferromagnetic exchange interaction
on the stability of the ferromagnetic phase with respect to dilution we
choose
\begin{equation}
\label{eq:J_R_longrange}
  J(R)=\begin{cases}
J^0\left(R/a\right)^{-4}&\text{for }a\leq R\leq \sqrt{n}a,\\
0& \text{otherwise.}
\end{cases}
\end{equation}
Here, $n$ is the number of the outermost shell with non-zero exchange
interaction (or the outermost shell that is taken into account
for the spin-spin interaction in the calculations), $a$ is again the
nearest-neighbor distance and $J^0$ the nearest-neighbor interaction
strength. For $n=1$, we recover the case of pure nearest-neighbor
interaction, which we discussed in the last subsection. For
$n\rightarrow\infty$, we have a long-range interaction with algebraic
fall-off. In this case, there is always only one cluster even for finite
dilution, and  we do not encounter complications due to smaller
clusters in the calculation of the Curie temperature.

\begin{figure}[tb]
\centerline{\includegraphics[width=1\linewidth]
{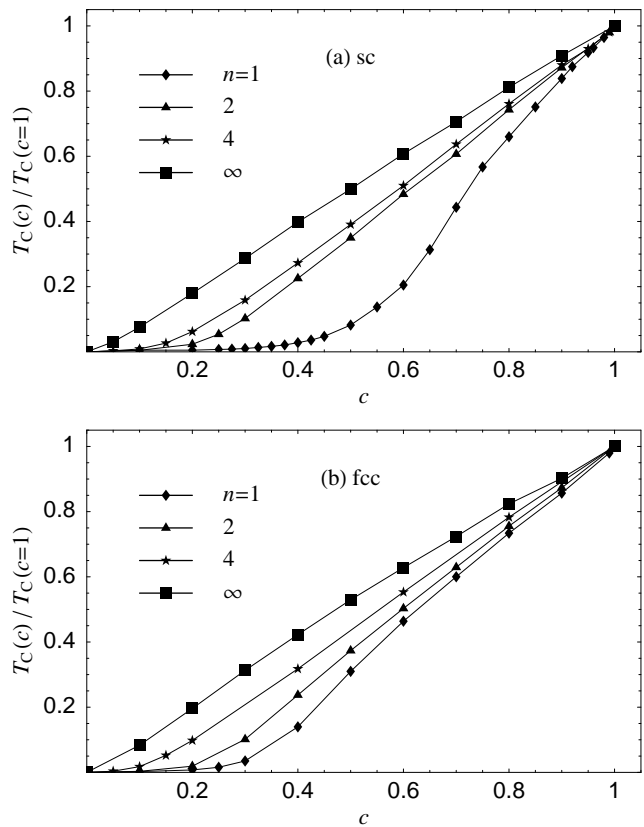}}
\caption{
\label{fig:longrange_Tc}
Reduced Curie temperature $\Tc(c)/\Tc(1)$ as function of the
concentration $c$ for exchange interactions with
different range $n$ for (a) sc and (b) fcc lattice structure
}
\end{figure}

In figure \ref{fig:longrange_Tc}, we plotted the Curie
temperature $\Tc$ as function of the concentration $c$ for different
ranges $n$ of the exchange interaction.
We performed calculations for the sc and fcc lattice structure, but no
qualitative differences could be found between them.
As expected, a decreasing value of the threshold concentration, above
one finds a finite Curie temperature, is found with increasing range $n$
of the interaction. Moreover, the observed values of the threshold
concentration agree quite well with 
the corresponding critical percolation concentration, which is a 
lattice-structure- and interaction-range-dependent
quantity.\cite{ElliottEtal60,Kirkpatrick73,Stauffer95}
Furthermore, the results show that the exchange interaction must have a
range beyond the fourth shell to obtain a finite
Curie temperature for concentration values of $c\approx0.05$, which are
typical of the recently studied DMS. This is consistent with the values
$n\geq5$ for the fcc and $n\geq6$ for the sc lattice obtained by
the formula
\begin{equation}
  n \gtrsim \left(\f{3}{4\pi c} f_{\text{geom}} \right)^{2/3}
\end{equation}
with $f_{\text{geom}}^{\text{sc}}=2.4$ and
  $f_{\text{geom}}^{\text{fcc}}=2.4/\sqrt{2}$, which is based on
  percolation arguments.\cite{LitvinovDugaev01}

For $n=\infty$, the Curie temperature is essentially linear in the
concentration for both lattice structures. Hence, a small, but
finite value is found for the Curie temperature even for very low
concentrations. Thus, it seems that the applicability of the
virtual-crystal approximation is restored in the limit
$n\to\infty$. However,
calculation for other forms of ferromagnetic long-range exchange
interactions show that the perfect linear dependence
$\Tc\propto c$ is a peculiarity of the exchange interaction
\eqref{eq:J_R_longrange}, although the proportionality seems to be roughly
fulfilled for all ferromagnetic long-range exchange
interactions.
Note that placing the remaining spins of the diluted system on an artificial
lattice with an appropriate lattice constant \mbox{$a(c)\propto c^{1/3}$} and
changing the exchange parameter according to
eq. \eqref{eq:J_R_longrange} gives a  Curie temperature \mbox{$\Tc\propto
c^{4/3}$} in contrast to the observed \mbox{$\Tc\propto c$}.

We also calculated magnetization curves for different
concentrations and exchange-interaction ranges for the sc and fcc lattice
structure. However, since any unusual features of the magnetization curves
have not been observed, we will refrain from an extensive discussion of
these.

\subsection{Oscillating Long-Range Interaction}
\label{subsec:Osc_Long_Range}

\begin{figure}[tb]
\centerline{\includegraphics[width=1\linewidth]
{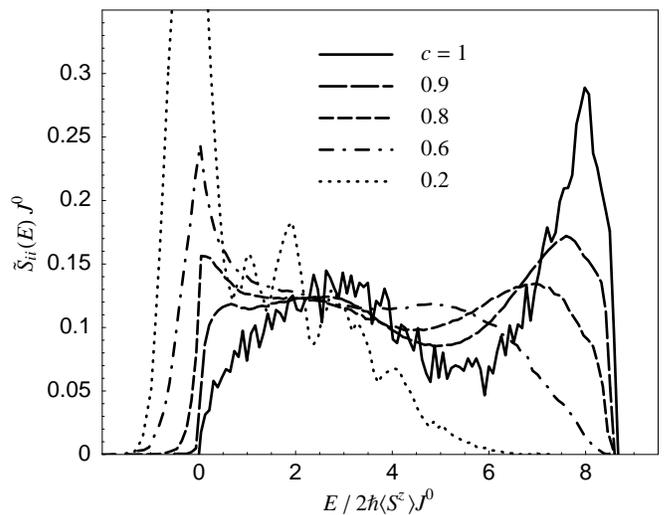}}
\caption{
\label{fig:osc_Sii}
  Site-averaged local magnon spectral density $\tilde{S}_{ii}(E)$
for oscillating long-range exchange interaction on an sc lattice for various
concentrations $c$ of spins (shown as histogram of the eigenvalue
distribution of the Hamilton  matrix, each curve is an average of 5 to
10 configurations of $8\,000$ to $14\,000$ spins)
}
\end{figure}

It is believed that the ferromagnetism in DMS is carrier-induced,
with holes in the valence band of the host semiconductor mediating the
interaction between the localized moments of the impurity atoms. One
feature of such an indirect exchange mechanism is that it may give rise to
a long-ranged effective exchange interaction  $J(R)$ which is
non-monotonous as function of the spin-spin distance $R$ and even may be
ferromagnetic for some, but antiferromagnetic for other values of $R$.
To study such an oscillatory exchange interaction, we choose
\begin{equation}
  \label{eq:Def_J_osc}
  J(R)=\begin{cases}
    J^0\left(\f{R}{a}\right)^{-3} 
    \cos\!\left(\, \pi\!\left(\f{R}{a}-1\right)\,\right) &
    \text{for } a\leq R,\\
    0& \text{otherwise.}\\
  \end{cases}
\end{equation}
In figure \ref{fig:osc_Sii}, we plotted the site-averaged local magnon
spectral density $\tilde{S}_{ii}(E)$ for different
concentrations. Although the exchange interaction \eqref{eq:Def_J_osc}
contains antiferromagnetic interactions, the concentrated spin system
has a saturated ferromagnetic ground state and a low-temperature
ferromagnetic phase with a Curie temperature of $\kB\Tc\approx1.6\hbar^2J^0$.
However, as soon as the concentration is decreased from $c=1$, the
spectral density becomes non-vanishing for $E<0$. This effect is still
small for high concentrations, and one could try to use the
computed eigenvalues to calculate Curie temperatures by using a small
shift of the lowest eigenvalues, but already at intermediate
concentrations the negative $\mbf{H}$-eigenvalues substantially
contribute to $\tilde{S}_{ii}$ and certainly can not be ignored any
more. This means that we can not compute any spontaneous magnetization
or finite Curie temperature for such systems with our method.

The results observed for the exchange interaction \eqref{eq:Def_J_osc} can
be generalized.
Eigenvalues $E_r<0$ of the Hamilton matrix $\mbf{H}$ occur for all
systems with $c<1$ and antiferromagnetic components in the exchange
interaction.
These negative eigenvalues stem from the fact that at finite dilution 
there are always some spins in the system which predominantly couple
antiferromagnetically to the other spins of the system. As a result, the
saturated ferromagnetic state is not the ground state of the diluted
system and hence, is thermodynamically unstable against creation of
magnons.
However, the Tyablikov approximation with the assumption of a
uniform magnetization is a reasonable approximation to the Heisenberg
model only for a saturated ferromagnetic ground state, and therefore, the
presented method fails for oscillatory exchange interactions.
Note that these considerations rule out a highly oscillating exchange
interaction for materials which are ferromagnetic
despite a low concentration of magnetic atoms.
For materials with highly oscillating exchange interaction, one rather
expects a spin-glass-like phase at low concentrations instead of a
ferromagnetic one.\cite{EggenkampEtal95}

\section{Summary}
\label{sec:Summary}

The aim of this article was to study the magnetic properties of diluted
spin systems. Our first step was to generalize the
Tyablikov approximation to spin systems without translational symmetry
to account for the substitutional disorder present in such systems. The
resulting equation of motion for the magnon Green's function  was
then solved numerically for finite spin systems. The obtained
spectral densities were then used to estimate the Curie temperature and the
spontaneous magnetization of the corresponding infinite system.

Compared with other methods\cite{Montgomery70,FooWu72,TK72,
  ElliottPepper73,KumarHarris73,HarrisEtal74,Nickel75,Kaneyoshi75,
  TheumannTK75,LageStinchcombe77,TK77,DveyAharonFibich78,
  TK83,KawasakiEtal92,KaneyoshiMascur93,ZhaoHuber96,BouzerarBruno02,
  KleninBlume76,HarrisKirkpatrick77,Klenin79,
  SchliemannEtal01,BerciuBhatt01,CalderonEtal02,  
  AlvarezEtal02,BreyGomezsantos03,BouzerarEtal03,
  Griffiths69,LitvinovDugaev01,KaminskiDasSarma03}, our approach has the
advantage that the spins are treated quantum-mechanically within the Tyablikov
approximation, which goes beyond mean-field and classical-spin approximation,
and -- apart from the simplification of a uniform magnetization --  no
approximations are needed 
with respect to the substitutional disorder. Furthermore, the numerical
effort as well as the analytical is fairly low in our approach.

Our calculations for short-range exchange interactions show no magnetic
order below the critical percolation concentration in accordance with
percolation theory.
For ferromagnetic long-range exchange interactions, the calculation show
a linear dependence of the Curie temperature on the concentration of
spins. Hence, a finite Curie temperature is found even for very low
concentrations.
For systems with oscillating long-range exchange interaction,
the presented method fails to give answers due to the lack of a
ferromagnetic ground state at finite dilution.
For ferromagnetic materials with concentrations of
magnetic atoms as low as $c\approx0.05$, these results imply that the effective
exchange interaction must be very long-ranged, but can not oscillate
strongly with the inter-spin distance.

In this article, we have shown that our
approach gives quite reasonable results for a variety of model systems
with ferromagnetic ground state. The next step should be an application
to real substances.
To attack the problem of ferromagnetic DMS, for example, the presented
method may be combined with a theory for the electronic degrees of
freedom, which provides the values for the effective exchange
interaction between the localized moments.
Furthermore, the method can be extended to account for a site-dependent
magnetization of the localized moments. However, this will highly
increase the numerical effort.
Clustering and other forms of chemical ordering may be easily
included into the model to make it more realistic for materials, in which
these play a role. Moreover, the method could then be used to analyze of
the effects of chemical short-range order on the magnetic properties.
This knowledge could be very useful for the development of new materials,
in which one can influence the distribution of the magnetic atoms by
choosing certain preparation parameters and hence, obtain certain
desired magnetic properties.
Furthermore, it should be investigated whether the presented method, in
particular the treatment of the zero-eigenvalues, could be put on a firm
theoretical basis.

%%%%%%%%%%%%%%%%%%%%%%%%%%%%%%%%%%%%%%%%%%%%%%%%%%%%%%%%%%%%%%%%%%%%%%

\begin{acknowledgments}

This work benefited from the support of the SFB290 of the Deutsche
Forschungsgemeinschaft.
One of us (S. Hilbert) gratefully acknowledges financial support by the
Friedrich-Naumann-Stiftung.
Thanks to C. Santos for helpful discussions.
\end{acknowledgments}

%%%%%%%%%%%%%%%%%%%%%%%%%%%%%%%%%%%%%%%%%%%%%%%%%%%%%%%%%%%%%%%%%%%%%%%%%%%

%\bibliography{literature}

\end{document}